\documentclass[12pt]{article}
\usepackage{epsf}
\usepackage{amsmath}
\usepackage{cite}
\usepackage{graphics}
\usepackage{amsfonts}
\usepackage{amssymb}
\usepackage{latexsym}
\usepackage{color}
\input{colordvi.tex}

\renewcommand{\thefootnote}{\#\arabic{footnote}}

\begin{document}
\setcounter{footnote}{0}

\begin{titlepage}

\begin{center}

\hfill HIP-2010-17/TH\\
\hfill BI-TP 2010/18\\

\vskip .5in

{\Large \bf Scale-dependence of Non-Gaussianity \\ in the Curvaton Model
}

\vskip .45in

{\large
Christian~T.~Byrnes$\,^{1}$, Kari~Enqvist$\,^{2,3}$ and Tomo~Takahashi$\,^4$
}

\vskip .45in

{\em
$^1$
Fakult\"{a}t f\"{u}r Physik, Universit\"{a}t Bielefeld, Postfach 100131, 33501 Bielefeld, Germany\\
$^2$
Helsinki Institute of Physics, University of Helsinki, PO Box 64, FIN-00014,
Finland \\
$^3$
Department of Physical Science, University of Helsinki, \\PO Box 64, FIN-00014,
Finland \\
$^4$
Department of Physics, Saga University, Saga 840-8502, Japan
}

\end{center}

\vskip .4in

\begin{abstract}

  We investigate the scale-dependence of $f_{\rm NL}$ in the self-interacting curvaton
  model. We show that the scale-dependence, encoded in the spectral index $n_{f_{\rm NL}}$, can be observable by
  future cosmic microwave background observations, such as CMBpol, in a significant
  part of the parameter space of the model. We point out that together
  with information about the trispectrum $g_{\rm NL}$, the self-interacting
  curvaton model parameters could be completely fixed by observations.
  We also discuss the scale-dependence of $g_{\rm NL}$ and its
  implications for the curvaton model, arguing that it could provide
  a complementary probe in cases where the theoretical value of $n_{f_{\rm NL}}$
  is below observational sensitivity.

\end{abstract}
\end{titlepage}

\renewcommand{\thepage}{\arabic{page}}
\setcounter{page}{1}
\renewcommand{\thefootnote}{\#\arabic{footnote}}

\section{Introduction}

It is now widely recognized that non-Gaussianity is a
useful tool to distinguish between various models for the primordial perturbation.
Although the power spectrum of the CMB temperature
fluctuations are well constrained by current observations such
as WMAP \cite{Komatsu:2010fb}, at linear order many mechanisms can give rise to almost
the same spectrum of fluctuations. Thus, to differentiate between models, one needs
to go beyond the linear order. For example, the standard slow-roll
inflation models predict an almost Gaussian primordial perturbation, while
other mechanism such as the curvaton
\cite{Enqvist:2001zp,Lyth:2001nq,Moroi:2001ct}, yield almost identical
power spectrum but can have a significant non-Gaussian component.

The deviations from the purely Gaussian perturbation
are usually characterized by some non-linearity parameters, such as
$f_{\rm NL}$, related to the 3-point function, and $\tau_{\rm NL}$ and $g_{\rm NL}$,
which are related to the 4-point function.  The present observational constraint on $f_{\rm NL}$ is
$-10 < f_{\rm NL} < 74 ~~{\rm (95 \%
  C.L.)}$ \cite{Komatsu:2010fb}\footnote{
For the constraints on $\tau_{\rm NL}$ and $g_{\rm NL}$, see \cite{Desjacques:2009jb,Vielva:2009jz,Smidt:2010sv} and for forecasts on future constraints see \cite{Kogo:2006kh,Smidt:2010ra}.
}.  If taken at face value, the current limit might actually suggest that $f_{\rm NL}$
deviates from zero (although it is still consistent with zero); this might spell trouble for the standard single-field slow-roll
inflation models, which typically give $f_{\rm NL}$ which is proportional to the small slow-roll parameters, or at
most $ f_{\rm NL}\sim
\mathcal{O}(1)$ \cite{Maldacena:2002vr}.
If this suggestion turns out to be confirmed by Planck, the curvaton
models would be natural candidates for the primordial perturbation.

Indeed, non-Gaussianity in the curvaton model has been the subject of extensive
study
\cite{Lyth:2002my,Dimopoulos:2003ss,Bartolo:2003jx,Lyth:2005fi,Enqvist:2005pg,Malik:2006pm,Sasaki:2006kq,Huang:2008ze,Ichikawa:2008iq,Multamaki:2008yv,Enqvist:2008gk,Huang:2008bg,Huang:2008zj,Kawasaki:2008mc,Chingangbam:2009xi,Enqvist:2009zf,Enqvist:2009eq,Enqvist:2009ww,Chingangbam:2010xn}.
However, in the majority of these studies
the  curvaton potential is assumed to be of the
quadratic form. Since in order to be able to decay, the curvaton has to couple to some other fields, it  then follows
that through loop corrections there will also appear some effective curvaton self-couplings, even if they were absent at tree level.
Moreover, it has been pointed out that even small deviations from the quadratic potential may have significant effects for the
primordial perturbation if due to the curvaton
\cite{Dimopoulos:2003ss,Enqvist:2005pg,Enqvist:2008gk,Huang:2008bg,Huang:2008zj,Kawasaki:2008mc,Chingangbam:2009xi,Enqvist:2009zf,Enqvist:2009eq,Enqvist:2009ww}.
In particular, it has been shown that for non-quadratic
curvaton potentials the predictions for
$f_{\rm NL}$ and $g_{\rm NL}$ can be quite different as compared with the purely
quadratic case \cite{Enqvist:2005pg,Enqvist:2008gk}. In the presence of non-linearities,  it is also possible that
the non-Gaussianity of the perturbation can be revealed by
$g_{\rm NL}$ (or the 4-point function) rather than by $f_{\rm NL}$, which by virtue of the non-linearites can remain very small
for certain choices of the parameters \cite{Enqvist:2008gk}.

Another point worth stressing is that the well-known relation
$f_{\rm NL} \sim 1 / r_{\rm dec}$ for the curvaton model, where
$r_{\rm dec}$ is related to the curvaton fraction of the total energy density at the time of curvaton decay,
 is not generic and strictly speaking only true for the quadratic potential \cite{Enqvist:2009ww}.
Thus a general, self-interacting curvaton model is a source for new
 observational signatures.  However, self-interaction means also more free parameters.  Hence it would be desirable to have more
observational quantities beyond $f_{\rm
  NL}$ and $g_{\rm NL}$ that would help to constrain the models.

Indeed, recently it has been pointed out that the scale-dependence of $f_{\rm
  NL}$ could be measured by future observations \cite{Sefusatti:2009xu}.
Theoretical aspects of the scale-dependence have been investigated in
\cite{Byrnes:2008zy,Kumar:2009ge,Byrnes:2009pe,Byrnes:2010ft}. We note that the scale dependence of equilateral type non-Gaussianity is also of interest \cite{Chen:2005fe,LoVerde:2007ri,Khoury:2008wj,RenauxPetel:2009sj}. Here we study the scale-dependence of the non-Gaussianity parameters
in a self-interacting
curvaton model. We show that, in some cases,
the scale-dependence of $f_{\rm NL}$ can be large enough to be
detected in future observations. We also demonstrate that by
using the combined information about $f_{\rm NL}$ and $g_{\rm NL}$, we may essentially
fix the curvaton model parameters.

The organization of this paper is as follows. In the next section, we
summarize the formalism for the study of the scale-dependence of
$f_{\rm NL}$, which manifests itself in the spectral index $n_{f_{\rm NL}}$, in a general
curvaton model. Then in Section~\ref{sec:results}, we closely look at
$n_{f_{\rm NL}}$ in various cases and discuss its detectability in
future observations. We also discuss the implications for $g_{\rm NL}$, in particular focusing on
the complementary nature of the combined information. The final section is devoted to the
summary and a discussion about the implications of the results.

\section{Formalism and Definitions}
\label{sec:formalism}

For the purpose of discussing the scale-dependence of non-Gaussianity
parameters in the curvaton model, let us assume that the potential for the curvaton $\sigma$ can be written as
\begin{equation}
\label{eq:V}
V(\sigma) = \frac{1}{2} m_\sigma^2 \sigma^2
+ \lambda m_\sigma^4  \left( \frac{\sigma}{m_\sigma} \right)^p,
\end{equation}
where $m_\sigma$ is the mass of the curvaton, and $\lambda$ and $p$ represent the
strength and the power of the self-coupling, respectively.  To characterize the relative size of the
self-interaction term in the potential, we define the parameter $s$ as
follows:
\begin{equation}
\label{eq:s}
s \equiv  2 \lambda \left( \frac{\sigma_\ast}{m_\sigma} \right)^{p-2},
\end{equation}
where the subscript $\ast$ represents that the quantity is
evaluated at the time the scale leaves the horizon. Thus $s$ is the ratio of two terms
in Eq.~\eqref{eq:V} at the horizon exit.  Once the curvaton begins to oscillate, the energy
density of the curvaton can soon be well
described by the purely quadratic case (i.e., the ``nearly"
quadratic case discussed in \cite{Enqvist:2005pg,Enqvist:2008gk}), the
curvature perturbation up to the third order, adopting sudden decay approximation, can be written as\footnote{
In the following, we do not consider isocurvature fluctuations.
For studies of isocurvature fluctuations in the curvaton model,
see e.g. \cite{Moroi:2002rd,Lyth:2003ip,beltran:2008ei,Moroi:2008nn,Takahashi:2009cx}.
}
\begin{eqnarray}
\label{eq:zeta_cur}
\zeta &=&
\frac{2}{3} r_{\rm dec} \frac{\sigma'_{\rm osc}}{\sigma_{\rm osc}}  \delta \sigma_\ast
+
\frac{1}{9} \left[ 3r_{\rm dec} \left(
1 +
\frac{\sigma_{\rm osc} \sigma_{\rm osc}^{\prime\prime}}{\sigma_{\rm osc}^{\prime 2}}
\right)
- 4 r_{\rm dec}^2 -2  r_{\rm dec}^3
\right]
\left( \frac{\sigma'_{\rm osc}}{\sigma_{\rm osc}} \right)^2  (\delta \sigma_\ast )^2 \notag \\
&&+
\frac{4}{81} \left[
\frac{9r_{\rm dec}}{4}  \left(
\frac{\sigma_{\rm osc}^2 \sigma_{\rm osc}^{\prime\prime\prime}}
{\sigma_{\rm osc}^{\prime 3}}
+
3\frac{\sigma_{\rm osc} \sigma_{\rm osc}^{\prime\prime}}{\sigma_{\rm osc}^{\prime 2}}
\right)
-9r_{\rm dec}^2
\left(
1
+
\frac{\sigma_{\rm osc} \sigma_{\rm osc}^{\prime\prime}}{\sigma_{\rm osc}^{\prime 2}}
\right)
\right. \notag \\
&&
\left.
+\frac{r_{\rm dec}^3}{2} \left(
1
-
9\frac{\sigma_{\rm osc} \sigma_{\rm osc}^{\prime\prime}}{\sigma_{\rm osc}^{\prime 2}}
\right)
+10r_{\rm dec}^4 + 3r_{\rm dec}^5
\right]
\left( \frac{\sigma'_{\rm osc}}{\sigma_{\rm osc}} \right)^3  (\delta \sigma_\ast )^3~,
\end{eqnarray}
where $r_{\rm dec}$ is defined as
\begin{equation}
r_{\rm dec} \equiv
\left.
\frac{3 \rho_\sigma}{4 \rho_{\rm rad} + 3\rho_\sigma}
\right|_{\rm decay}.
\end{equation}
We assume that, for a given inflationary Hubble parameter $H_*$,  the curvaton decay rate can be adjusted such that the correct perturbation amplitude $\zeta\simeq 10^{-5}$ can be obtained.

The non-linearity parameter  $f_{\rm NL}$ is then given
by\footnote{
For the definition and details of the calculations of the non-linearity parameters
in the model, we refer the reader to e.g., \cite{Enqvist:2008gk}.
}
\begin{equation}
\label{eq:fNL_cur}
f_{\rm NL} = \frac{5}{4r_{\rm dec}} \left(
1 +
\frac{\sigma_{\rm osc} \sigma_{\rm osc}^{\prime\prime}}{\sigma_{\rm osc}^{\prime 2}}
\right)
-\frac{5}{3} - \frac{5r_{\rm dec}}{6}.
\end{equation}

Let us now discuss the scale dependence of $f_{\rm NL}$ in the self-interacting curvaton
model.  The scale dependence of $f_{\rm NL}$
is defined as (see \cite{Byrnes:2009pe} for details)
\begin{equation}
\label{eq:def_nfnl}
n_{f_{\rm NL}} \equiv \frac{ d \ln | f_{\rm NL} |}{d \ln k},
\end{equation}
where we regard $n_{f_{\rm NL}}$ as the spectral index for $f_{\rm
  NL}$.  From this definition, treating $n_{f_{\rm NL}}$ as a constant is equivalent to assuming the power-law form:
\begin{equation}
\label{eq:power_law}
f_{\rm NL} \propto k^{n_{f_{\rm NL}}}.
\end{equation}

For the self-interacting curvaton model, the spectral index can be
calculated as \cite{Byrnes:2009pe} (for a very recent, alternative method of calculation see \cite{Byrnes:2010ft})
\begin{eqnarray}
\label{eq:nfNL}
n_{f_{\rm NL}}
&=&
\frac{V^{\prime\prime\prime} (\sigma_\ast)}{3 H_\ast^2}
\frac{\sigma_{\rm osc}}{4\sigma'_{\rm osc}}
\left[ \frac{1}{4}  \left(
1 +
\frac{\sigma_{\rm osc} \sigma_{\rm osc}^{\prime\prime}}{\sigma_{\rm osc}^{\prime 2}}
\right) - \frac{1}{3} r_{\rm dec} -  \frac{1}{6}r_{\rm dec}^2 \right]^{-1}  \notag \\
&=&
\frac{V^{\prime\prime\prime} (\sigma_\ast)}{3 H_\ast^2}
\frac{\sigma_{\rm osc}}{\sigma'_{\rm osc}}
\frac{5}{4 r_{\rm dec} f_{\rm NL}},
\end{eqnarray}
where a prime indicates the derivative with respect to $\sigma_\ast$,
i.e., $V^{'''} = d^3 V / d\sigma^3$.

In addition to the spectral index $n_{f_{\rm NL}}$,
the running of $n_{f_{\rm NL}}$ can also be defined analogously to the running of the spectral index of the usual
spectrum as
\begin{equation}
\alpha_{f_{\rm NL}} \equiv \frac{d n_{f_{\rm NL}}}{d \ln k}.
\end{equation}
For the curvaton model with self-interaction, we obtain the following
expression for the running $\alpha_{f_{\rm NL}}$\footnote{
We thank Qing-Guo Huang for pointing out an error in this
formula in previous versions of this paper. The result given here agrees
with the arXiv version of \cite{Huang:2010cy}.
}:
\begin{eqnarray}
\label{eq:alpha_fNL}
\alpha_{f_{\rm NL}} &=&
n_{f_{\rm NL}}
\left[ - \frac{ V^{\prime\prime}}{3H_\ast^2}
+2 \epsilon - \frac{V^{\prime\prime\prime\prime}}{3H_\ast^2}
\frac{ V^{\prime}}{V^{\prime\prime\prime} }
\right] - n_{f_{\rm NL}}^2.
\end{eqnarray}

The derivative of the potential can be evaluated once $s, p, \lambda$,
and the mass $m_\sigma$ are given. However, in the following
discussion, instead of the mass we find it convenient to use the  parameter $\eta_{\sigma\sigma}$,
defined as the ratio between $V^{\prime\prime}$ and the
Hubble parameter during inflation:
\begin{equation}
\label{defeta}
\eta_{\sigma\sigma}
= \frac{V^{''}(\sigma_\ast)}{3H_\ast^2}
= \frac{m_\sigma^2}{3H_\ast^2} \left( 1+ \frac{p(p-1)}{2} s \right).
\end{equation}
Notice that the spectral index for the power spectrum in this model is
given by
\begin{equation}
n_s - 1 = - 2 \epsilon + 2 \eta_{\sigma\sigma},
\end{equation}
where $\epsilon$ is the slow-roll parameter for the inflaton sector.
Thus the value of $\eta_{\sigma\sigma}$ should be small enough to be
consistent with the observational constraint: $ 0.9358 < n_s -1 <
0.9921$ (95\% C.L.) \cite{Komatsu:2010fb}.

With the definition (\ref{defeta}) for  $\eta_{\sigma\sigma}$, we can express the higher derivatives
of the potential as
\begin{eqnarray}
\frac{V^{'''}}{3H_\ast^2} & = &
\eta_{\sigma\sigma}  \frac{p(p-1)(p-2) s}{2 + p(p-1) s} \frac{1}{\sigma_\ast},
\\
\frac{V^{''''}}{3H_\ast^2} & = &
\eta_{\sigma\sigma}  \frac{p(p-1)(p-2) (p-3)s}{2 + p(p-1) s} \frac{1}{\sigma_\ast^2}.
\end{eqnarray}

It should be noticed here that, in the self-interacting curvaton
model, the value of $f_{\rm NL}$ can change its sign when
parameters are varied. In fact, when $f_{\rm NL}$ changes its sign,
the scale-dependence of $f_{\rm NL}$ cannot be well described by the
power-law form of Eq.~\eqref{eq:power_law}.  One could also see this
by noticing that the expression of $n_{f_{\rm NL}}$ given by
Eq.~\eqref{eq:nfNL} diverges when $f_{\rm NL}=0$. However, as we will
discuss later, when $f_{\rm NL}$ is small enough, it would be impossible to
measure the scale-dependence of $f_{\rm NL}$; for small enough $f_{\rm NL}$ , it is not an observable.  Furthermore, we have checked
that  the magnitude of the running $\alpha_{f_{\rm NL}}$ is much smaller than
$n_{f_{\rm NL}}$ except in the region where $f_{\rm NL} \lesssim
\mathcal{O}(1)$.  In the following we do not consider
the unobservable region $f_{\rm NL} \lesssim \mathcal{O}(1)$; as a consequence, for us $n_{f_{\rm
    NL}}$ is indeed the proper way to describe scale-dependence of $f_{\rm NL}$ in the whole range of interest.

Although in
this paper we mainly focus on the scale-dependence of $f_{\rm NL}$ ,
we can also define the scale dependence of $g_{\rm NL}$
similarly to the case of $f_{\rm NL}$ as \cite{Byrnes:2010ft}
\begin{equation}
n_{g_{\rm NL}} \equiv \frac{ d \ln | g_{\rm NL} |}{d \ln k}.
\end{equation}
Since in the curvaton model with a self-interaction term, the non-linearity
parameter $g_{\rm NL}$ is given by
\begin{eqnarray}
\label{eq:gNL_cur}
g_{\rm NL} &=& \frac{25}{54}
\Big[
\frac{9}{4r_{\rm dec}^2}  \left(
\frac{\sigma_{\rm osc}^2 \sigma_{\rm osc}^{\prime\prime\prime}}
{\sigma_{\rm osc}^{\prime 3}}
+
3\frac{\sigma_{\rm osc} \sigma_{\rm osc}^{\prime\prime}}{\sigma_{\rm osc}^{\prime 2}}
\right)
-\frac{9}{r_{\rm dec}}
\left(
1
+
\frac{\sigma_{\rm osc} \sigma_{\rm osc}^{\prime\prime}}{\sigma_{\rm osc}^{\prime 2}}
\right)
\nonumber\\
&+&\frac{1}{2} \left(
1
-
9\frac{\sigma_{\rm osc} \sigma_{\rm osc}^{\prime\prime}}{\sigma_{\rm osc}^{\prime 2}}
\right)
+10r_{\rm dec} + 3r_{\rm dec}^2
\Big],
\end{eqnarray}
as in the case for $n_{f_{\rm NL}}$, a similar calculation (consistent with the results of
\cite{Byrnes:2010ft})  yields
$n_{g_{\rm NL}}$:
\begin{eqnarray}
&&n_{g_{\rm NL}}  =
\left[
 \frac{9}{4} \left( \frac{V^{''''}(\sigma_\ast)}{3H_\ast^2} -3\eta_{\sigma\sigma}^2 \right)
\left( \frac{\sigma_{\rm osc}}{\sigma_{\rm osc}'} \right)^2
  +
  \frac{27}{4}\left( \frac{V^{'''}(\sigma_\ast)}{3H_\ast^2} \right)
  \left\{
  \frac{\sigma_{\rm osc}^2 \sigma_{\rm osc}^{''}}{(\sigma_{\rm osc}')^3}
  +\left( 1 - \frac{4}{3} r_{\rm dec}  - \frac{2}{3} r_{\rm dec}^2 \right)
  \frac{\sigma_{\rm osc}}{\sigma_{\rm osc}'}
  \right\}
  \right] \notag \\
 &&
 \times \left[ \frac{9}{4}
 \left(
\frac{\sigma_{\rm osc}^2 \sigma_{\rm osc}^{\prime\prime\prime}}
{\sigma_{\rm osc}^{\prime 3}}
+
3\frac{\sigma_{\rm osc} \sigma_{\rm osc}^{\prime\prime}}{\sigma_{\rm osc}^{\prime 2}}
\right)
-9r_{\rm dec}
\left(
1
+
\frac{\sigma_{\rm osc} \sigma_{\rm osc}^{\prime\prime}}{\sigma_{\rm osc}^{\prime 2}}
\right)
+ \frac{r_{\rm dec}^2}{2}
\left(
1
- 9
\frac{\sigma_{\rm osc} \sigma_{\rm osc}^{\prime\prime}}{\sigma_{\rm osc}^{\prime 2}}
\right)
+ 10 r_{\rm dec}^3 + 3 r_{\rm dec}^4
\right]^{-1}. \notag \\
\end{eqnarray}

\section{$n_{f_{\rm NL}}$ in the curvaton model}
\label{sec:results}

\begin{figure}[htbp]
  \begin{center}
    \resizebox{160mm}{!}{
\hspace{-2cm}   \includegraphics{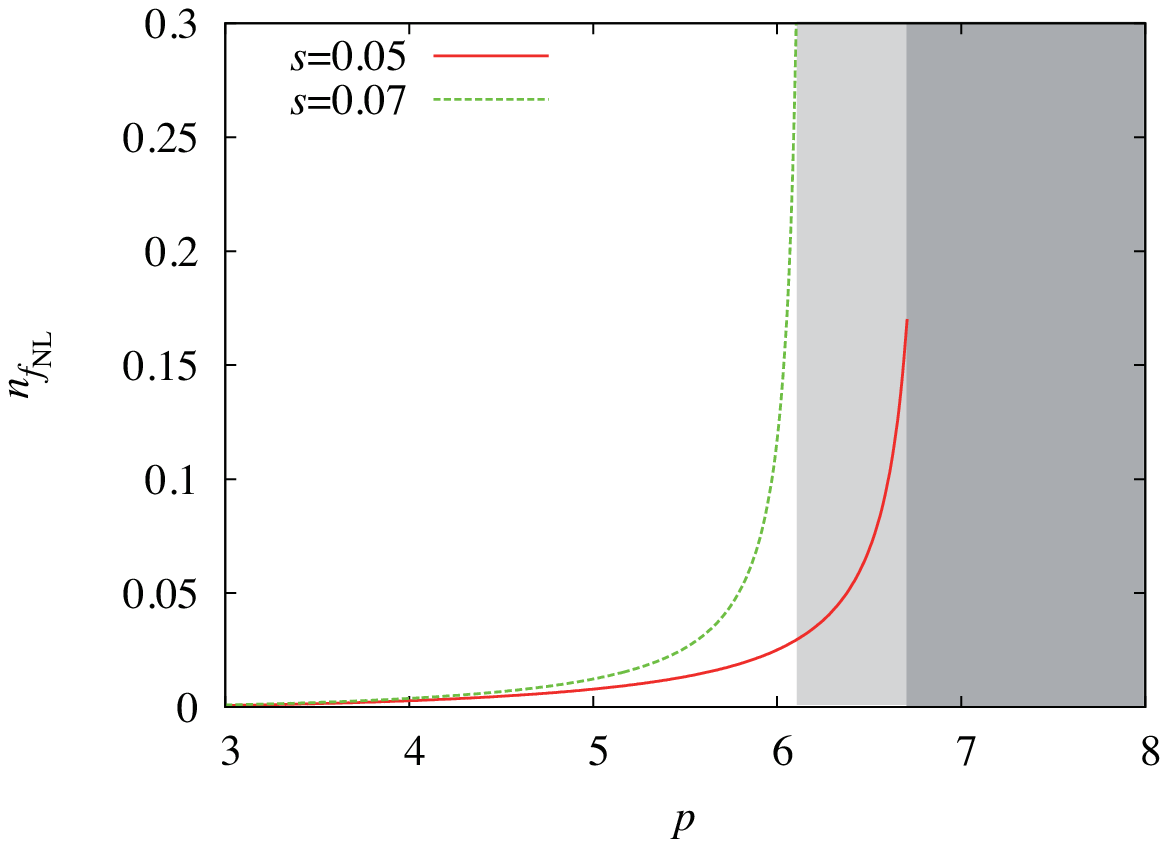}
  \includegraphics{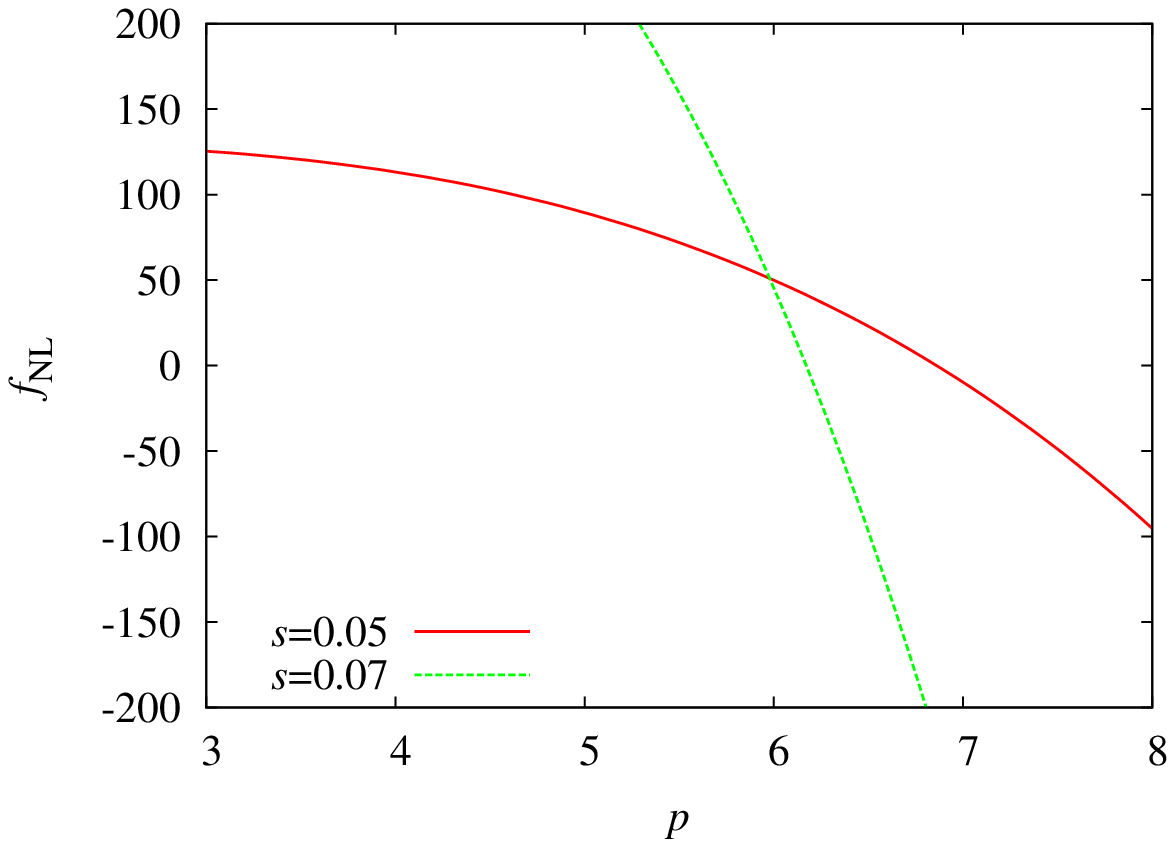}
}
  \end{center}
  \caption{Plots of $n_{f_{\rm NL}}$ (left panel) and $f_{\rm NL}$
    (right panel) as a function of the power $p$.  Here we take
    $s=0.05$ (red line) and $0.07$ (green line) with
    $\eta_{\sigma\sigma}=0.005$. The shaded areas correspond
    to the regions where the
power-law form $ f_{\rm
      NL} \propto k^{n_{f_{\rm NL}}}$ does not hold.  }
  \label{fig:nfNL}
\end{figure}

We have scanned the parameter space and evaluated the spectral index $n_{f_{\rm NL}}$ numerically.  In the left panel of
Fig.~\ref{fig:nfNL}, we plot the value of $n_{f_{\rm NL}}$ as a function of
the self-interaction power $p$ .  In the figure, we have chosen the parameters as
$s=0.05$ (red solid line), $0.07$ (green dashed line) and for the rest of this paper we choose
\begin{equation} \eta_{\sigma\sigma} =0.005.  \end{equation}
Note that the scale dependence of both $n_{f_{\rm NL}}$ and $n_{g_{\rm NL}}$ is proportional to $\eta_{\sigma\sigma}$, so they would both be larger if we chose a larger $\eta_{\sigma\sigma}$.
The corresponding values of
$f_{\rm NL}$ are also shown in the right panel.  We have chosen the
value of $r_{\rm dec}$ such that we obtain  $f_{\rm NL}=50$ at $p=6$ for
each relative self-interaction strength $s$ (therefore the values of $r_{\rm dec}$ are different for each case).
As discussed in the previous section, when $f_{\rm NL}$ very small, with
$f_{\rm NL} \lesssim \mathcal{O}(1)$, the power-law form cannot well
describe the scale dependence of $f_{\rm NL}$. To emphasize this fact we cut
out the regions with $f_{\rm NL} \le 10$.

\begin{figure}[htbp]
  \begin{center}
    \resizebox{160mm}{!}{
 \includegraphics{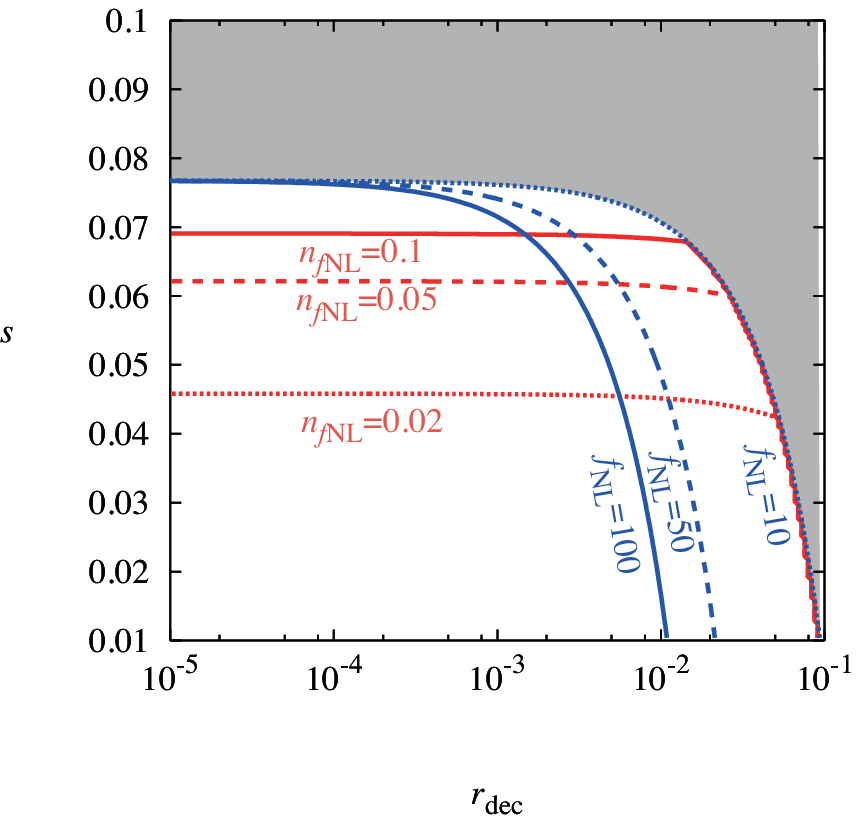}
\hspace{0.5cm}
  \includegraphics{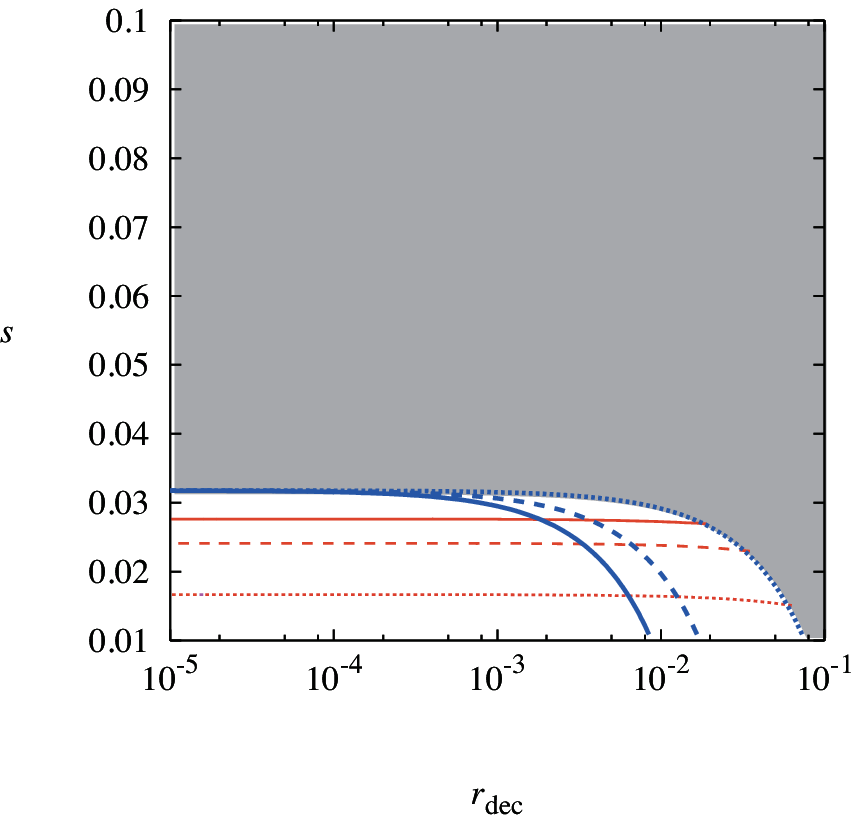}
}
  \end{center}
  \caption{Contours of $f_{\rm NL}$ and $n_{f_{\rm NL}}$ for $p=6$
    (left panel) and $8$ (right panel). The region with $f_{\rm NL} <
    10 $ is shaded. }
  \label{fig:n6_8}
\end{figure}

\begin{figure}[htbp]
  \begin{center}
    \resizebox{150mm}{!}{
    \includegraphics{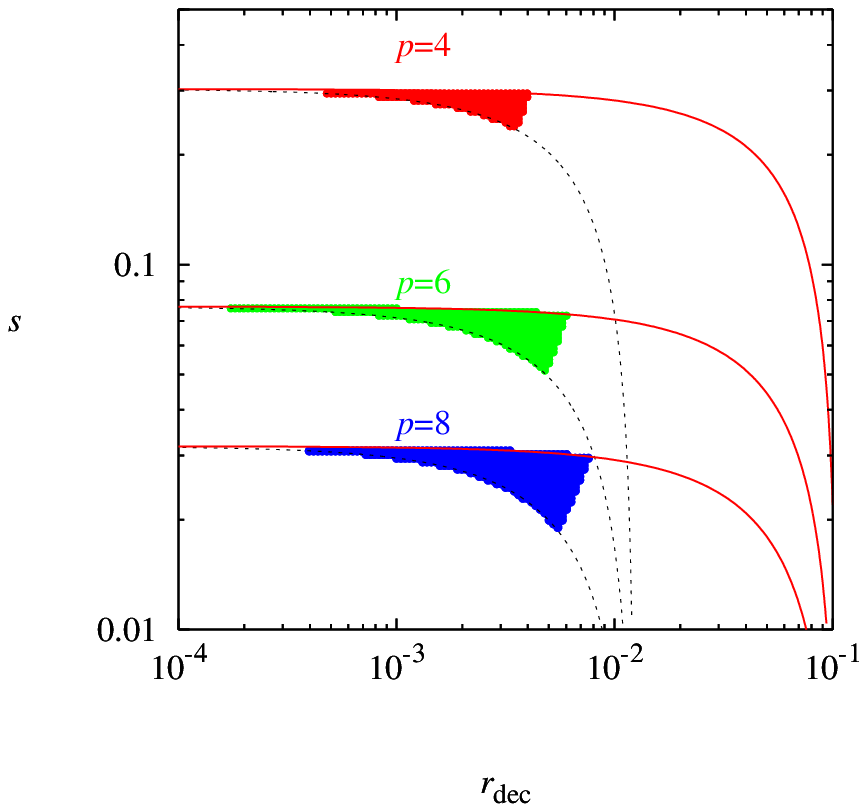}
}
  \end{center}
  \caption{Regions testable in future observations.  Shown is the
    parameter space where the theoretical prediction for $n_{f_{\rm
        NL}}$ exceeds $\Delta n_{f_{\rm NL}}$ given in
    Eq.~\eqref{eq:nfNL_obs}, (i.e., region where $n_{f_{\rm NL}} >
    \Delta n_{f_{\rm NL}} = 0.05 \times (50/f_{\rm NL})$ is
    satisfied), for $ 10 \le f_{\rm NL} \le 100$.  }
  \label{fig:testable}
\end{figure}

According to \cite{Sefusatti:2009xu}, future observations such as
CMBpol may be able to detect the scale dependence of $f_{\rm NL}$.
The observational error has been estimated as (for the fiducial value of
$n_{f_{\rm NL}}=0$)
\begin{equation}
\label{eq:nfNL_obs}
\Delta n_{f_{\rm NL}} \simeq 0.05 \frac{50}{f_{\rm NL}} \frac{1}{\sqrt{f_{\rm sky}}},
\end{equation}
where $f_{\rm sky}$ is the fraction of the sky measured. Hence in the parameter space where the theoretical estimate $n_{f_{\rm NL}}\ge \Delta n_{f_{\rm NL}}$
we may be able to constrain the
self-interacting curvaton model by the scale dependence of $f_{\rm
  NL}$. For example, as can be read off from figure \ref{fig:n6_8}, when $p=6$
and $s=0.07$, the spectral index is
$n_{f_{\rm NL}} \simeq 0.1$, which should be observable with future observations.

To chart the parameter space where $n_{f_{\rm NL}}$ is a useful tool for
model building,   we plot in Fig.~\ref{fig:n6_8} the contours of $n_{f_{\rm NL}}$ and
$f_{\rm NL}$ in the $s$--$r_{\rm dec}$ plane for the cases of $p=6$
and $8$.  From the expression of Eq.~\eqref{eq:fNL_cur}, one can see
that while generally $f_{\rm NL}$ is proportional to $1/r_{\rm dec}$, in
the self-interacting curvaton model $f_{\rm NL}$ also depends on $s$ through
the non-linear evolution of the curvaton field. This is encoded in
$\sigma_{\rm osc}$ and its derivatives.  On the other hand, when
$r_{\rm dec}$ is small, $n_{f_{\rm NL}}$ does not depend on $r_{\rm
  dec}$ but strongly depends on $s$. Since the scale-dependence of
$f_{\rm NL}$ mainly originates from the non-linear evolution of
$\sigma$ due to the higher order term in the potential, the value of $s$
controls $n_{f_{\rm NL}}$.  Notice that, when $r_{\rm dec} \ll \mathcal{O}(1)$,
the sign of $f_{\rm NL}$ is determined by the factor $ 1 + \sigma_{\rm
  osc} \sigma_{\rm osc}^{\prime\prime} /\sigma_{\rm osc}^{\prime 2}$
which appears in the first term in Eq.~\eqref{eq:fNL_cur}. This factor
depends on $s$ and $p$, but not on $r_{\rm dec}$.  This is the reason
why the contours of $f_{\rm NL}$ in Fig.~\ref{fig:n6_8} appear to
converge at $s \simeq 0.077$ and $0.032$ for $p=6$ and $8$,
respectively, where $f_{\rm NL}$ abruptly switches from a positive value
to a negative one. The region around this line also corresponds to the breakdown of
the power-law description for the scale-dependence.  By
cutting out the region with $f_{\rm NL} < 10$, which anyway is irrelevant from an
observational viewpoint, we therefore guarantee that
the power-law description is valid.

Since the sensitivity of $n_{f_{\rm NL}}$ in future observations
depends on the value of $f_{\rm NL}$, as indicated in
Eq.~\eqref{eq:nfNL_obs}, the self-interacting curvaton model may be tested in certain regions of the parameter space.
This is depicted in
Fig.~\ref{fig:testable}, where the regions with $n_{f_{\rm
    NL}} \ge \Delta n_{f_{\rm NL}}$ are shown for the cases $p=4,
6$ and $8$.  Here we conservatively adopt the range
$10 < f_{\rm NL} < 100$.  If future
observations find a non-zero value for $n_{f_{\rm NL}}$, the self-interacting curvaton
model must lie inside the colored regions. On the other hand, if no scale-dependence of $f_{\rm NL}$ will be detected, the colored region
will be excluded. Planck is expected to have about half of the sensitivity to $n_{f_{\rm NL}}$ compared to (\ref{eq:nfNL_obs}), but large scale structure probes may become even more sensitive to $n_{f_{\rm NL}}$ than CMBpol.

\section{Constraints implied by the trispectrum: $g_{\rm NL}$}

%
\begin{figure}[htbp]
  \begin{center}
    \resizebox{180mm}{!}{
\hspace{-2cm}   \includegraphics{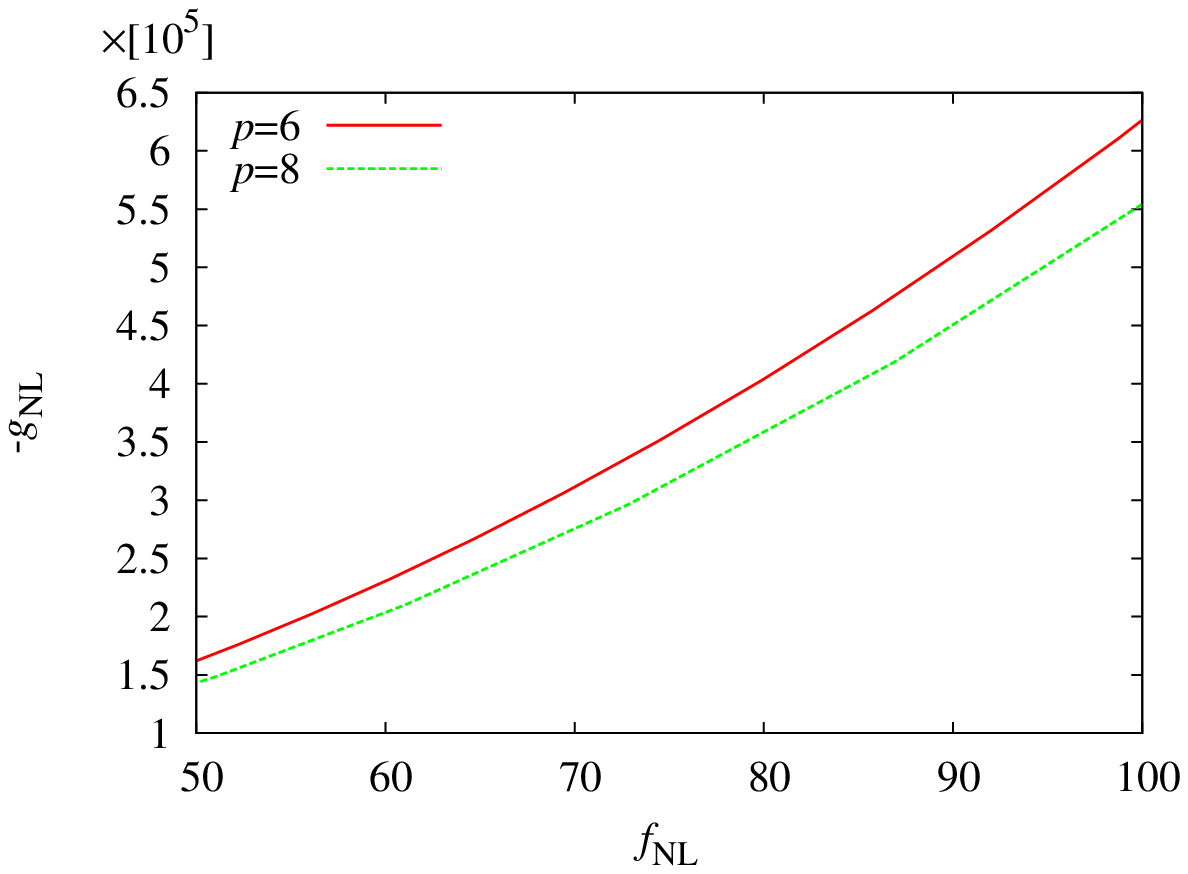}
  \includegraphics{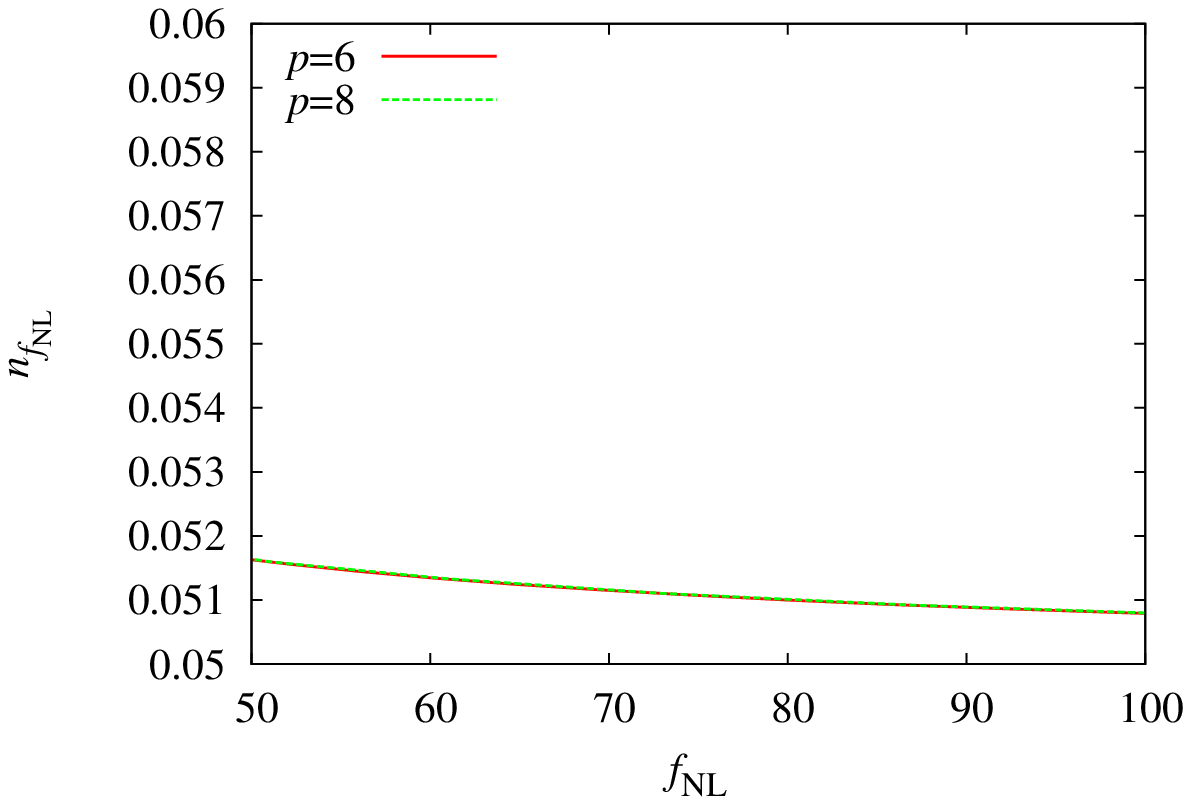}
}
  \end{center}
  \caption{Plots of $n_{f_{\rm NL}}$ (right) and $g_{\rm NL}$ (left)
    for the cases with $(p,s)=(6,0.0622)$ and $(8, 0.0214)$. The value
    of $s$ is chosen such that the cases of $p=6$ and $8$ give almost
    the same values of $n_{f_{\rm NL}}$ as a function of $f_{\rm
      NL}$. }
  \label{fig:gNL}
\end{figure}

\begin{figure}[htbp]
  \begin{center}
    \resizebox{100mm}{!}{
    \includegraphics{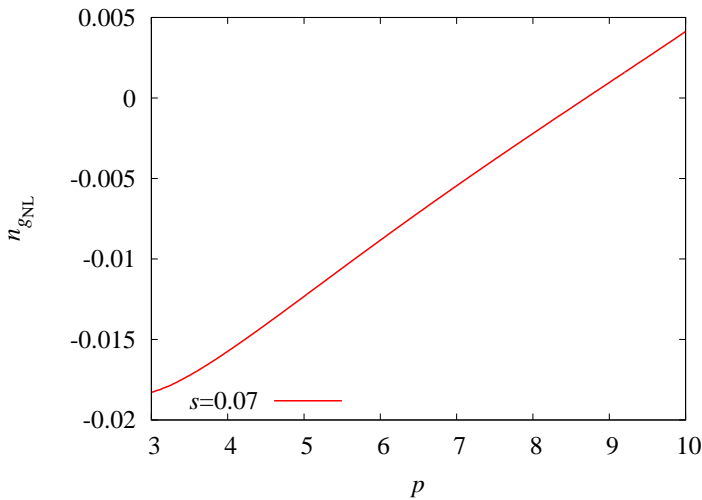}
}
  \end{center}
  \caption{Plot of $n_{g_{\rm NL}}$ as a function of the power $p$ for
    $s=0.07$.  }
  \label{fig:ngNL}
\end{figure}

In the previous section, we have shown that $n_{f_{\rm NL}}$ can be a
useful tool to probe/test the self-interacting curvaton model.
However, even if we can observe $n_{f_{\rm NL}}$ and $f_{\rm NL}$,
the situation remains degenerate:
different combinations of the power $p$, the strength $s$ and $r_{\rm dec}$ can give
the same values of $f_{\rm NL}$ and $n_{f_{\rm NL}}$.  To demonstrate this, in
the right panel of Fig.~\ref{fig:gNL}, we plot $n_{f_{\rm
    NL}}$ as a function of $f_{\rm NL}$, which is almost equivalent to
the plot as a function of $r_{\rm dec}$, for the cases of $p=6$ and
$8$.  To give an example, by fixing $s=0.0622$
and $0.0214$ for $p=6$ and $8$, respectively, the lines for $p=6$ and
$8$ almost coincide, as is shown in the right panel of Fig.~\ref{fig:gNL}.
However, the predictions for $g_{\rm NL}$ are  in general
different. This is demonstrated in the left panel of Fig.~\ref{fig:gNL}, where we show
$g_{\rm NL}$ as a function of $f_{\rm NL}$ for the same parameters as
in the right panel.  This figure indicates that, in principle,
it is possible break the degeneracy between $n_{f_{\rm NL}}$ and $f_{\rm NL}$
by including $g_{\rm NL}$ in the analysis. Thus combining together information
about the bispectrum amplitude, the spectral index of the bispectrum, and the amplitude
of the trispectrum, one may actually fix completely the parameters of
the simplest self-interacting curvaton model (\ref{eq:V}). We note that $\tau_{\rm NL}$ in this model does not give new information, since it satisfies $\tau_{\rm NL}=(6f_{\rm NL}/5)^2$ (for more discussion on the trispectrum see \cite{Seery:2006js,Byrnes:2006vq}), but testing this consistency relation is still valuable since it can determine whether it was justified to neglect the the inflaton field fluctuations \cite{Ichikawa:2008iq}.

Furthermore, we can also address the scale-dependence of $g_{\rm
  NL}$; the appropriate expression was already given in Sec.~\ref{sec:formalism}. In
Fig.~\ref{fig:ngNL}, $n_{g_{\rm NL}}$ is plotted as a function of the
power $p$ for $s=0.07$.  It is worth noticing that $n_{g_{\rm NL}}$
becomes larger the smaller $p$ is.  This is in contrast to the behavior of
$n_{f_{\rm NL}}$, which becomes smaller as the value of the
power $p$ becomes smaller, as is shown in Fig.~\ref{fig:nfNL}.  Thus,  in future observations
it may be possible to see
the scale-dependence of $g_{\rm NL}$ even if we cannot detect the scale-dependence of
$f_{\rm NL}$.  Hence, in addition to $g_{\rm NL}$
itself, the scale-dependence of $g_{\rm NL}$ might also provide a
tool to probe/test the self-interacting curvaton model.

\section{Conclusion and discussion}

In this paper, we have discussed the scale dependence of the
non-linear parameters such as $f_{\rm NL}$ and $g_{\rm NL}$ in a
curvaton model with a simple self-interaction term given by the
potential (\ref{eq:V}). The free parameters of the model are the
curvaton mass $m$, the power of the self-interaction term $p$, and
its relative strength $s$. 
The complete parametrization of the curvature perturbation would require also the knowledge of $H_*$, determined by the inflaton sector. 
In principle $H_*$ can be observationally fixed by a detection of
primordial gravitational waves, but their amplitude is likely to be very
small in the curvaton scenario. 
In a complete model, the curvaton decay rate might not be a totally independent parameter, but here we assume that, for a given $H_*$, the decay rate can be tuned to yield the correct perturbation amplitude $\zeta\simeq 10^{-5}$. In such a situation, as has been argued before,
 even a modest
amount of non-linearity in the curvaton potential \cite{Enqvist:2005pg,Enqvist:2008gk}
may lead to observable consequences. Therefore, although $f_{\rm
NL}$ is scale-invariant when the curvaton potential is quadratic, in
many cases the spectral index  $n_{f_{\rm NL}}$ of the bispectrum
can be detectable when one includes the self-interaction term; this
fact is depicted in Fig.~\ref{fig:nfNL}.

Since the observational sensitivity to the scale-dependence of
$f_{\rm NL}$ depends on the magnitude of $f_{\rm NL}$, there are regions where the
self-interacting curvaton model parameters can be pinpointed
by the observations of $f_{\rm NL}$ and its scale dependence $n_{f_{\rm NL}}$. The
testable domains in the parameter space are shown in Fig.~\ref{fig:testable}.
If no scale-dependence of $f_{\rm NL}$ will be found, these
regions would be excluded so that for the self-interacting curvaton model, the scale-dependence
can indeed provide us with clear-cut evidence for or against the model.

In addition to the scale-dependence of $f_{\rm NL}$, we also
studied the non-linearity parameter of the trispectrum, $g_{\rm NL}$,
and its scale-dependence $n_{g_{\rm NL}}$.  As is shown in
Fig.~\ref{fig:gNL}, even though  in principle there is a degeneracy in the
space of model parameters in that a given pair of values for
$n_{f_{\rm NL}}$ and $f_{\rm NL}$ is produced by several combinations of $(p,s)$, the
degeneracy can be broken by $g_{\rm NL}$. The spectral index of $g_{\rm NL}$ could
be used to probe the scale-dependence of
non-Gaussianity in the self-interacting curvaton model at least for some of the cases where
$n_{f_{\rm NL}}$ is not observable.

So far the studies of non-Gaussianity in the curvaton and other
models have focused on the non-linearity parameters $f_{\rm NL}, \tau_{\rm
  NL}$ and $g_{\rm NL}$. Here we have pointed out that the scale dependence could also
  provide a realistic and useful tool to test the models. In particular, we have argued
  that the self-interacting curvaton model parameters could actually be completely fixed
  by the combined observations of $f_{\rm NL}$, $g_{\rm NL}$, and the spectral
  index $n_{f_{\rm NL}}$. The fact that the projected observational sensitivity
  of the latter appears in many cases to be much higher than would be required
  for verification of the theoretical estimates is of course encouraging. It
  also lends some hope for the expectation that the physical origin of the
  primordial perturbation can be determined in the not too distant future.

\bigskip
\bigskip

\noindent {\bf Acknowledgments:} 
We thank Sami Nurmi for useful comments on the draft.
We also thank Qing-Guo Huang for pointing out an error in Eq.~\eqref{eq:alpha_fNL}.
T.T. would like to thank the Helsinki Institute of Physics for the hospitality during the visit,
where a part of this work has been done.  This work is supported in
part the Grant-in-Aid for Scientific Research from the Ministry of
Education, Science, Sports, and Culture of Japan No.\,19740145
(T.T.), and in part by the Academy of Finland grants 218322 and
131454 (K.E.). The authors also wish to thank
the organizers of the workshops ``The non-Gaussian universe" (YITP-T-09-05) and
``Cosmology -- The Next Generation" (YKIS2010 symposium), both at the Yukawa Institute for Theoretical Physics, Kyoto, Japan,
for hospitality, where parts of this work were done.


\begin{thebibliography}{100}

\bibitem{Komatsu:2010fb}
  E.~Komatsu {\it et al.},
  arXiv:1001.4538 [astro-ph.CO].


\bibitem{Enqvist:2001zp}
K.~Enqvist and M.~S.~Sloth,
Nucl.\ Phys.\ B {\bf 626}, 395 (2002)
[arXiv:hep-ph/0109214];

\bibitem{Lyth:2001nq}
D.~H.~Lyth and D.~Wands,
Phys.\ Lett.\ B {\bf 524}, 5 (2002)
[arXiv:hep-ph/0110002];

\bibitem{Moroi:2001ct}
T.~Moroi and T.~Takahashi,
Phys.\ Lett.\ B {\bf 522}, 215 (2001)
[Erratum-ibid.\ B {\bf 539}, 303 (2002)]
[arXiv:hep-ph/0110096].


\bibitem{Desjacques:2009jb}
  V.~Desjacques and U.~Seljak,
  Phys.\ Rev.\  D {\bf 81}, 023006 (2010)
  [arXiv:0907.2257 [astro-ph.CO]].

\bibitem{Vielva:2009jz}
  P.~Vielva and J.~L.~Sanz,
  arXiv:0910.3196 [astro-ph.CO].

\bibitem{Smidt:2010sv}
  J.~Smidt, A.~Amblard, A.~Cooray, A.~Heavens, D.~Munshi and P.~Serra,
  arXiv:1001.5026 [astro-ph.CO].

\bibitem{Kogo:2006kh}
  N.~Kogo and E.~Komatsu,
  Phys.\ Rev.\  D {\bf 73}, 083007 (2006)
  [arXiv:astro-ph/0602099].

\bibitem{Smidt:2010ra}
  J.~Smidt, A.~Amblard, C.~T.~Byrnes, A.~Cooray, A.~Heavens and D.~Munshi,
  Phys.\ Rev.\  D {\bf 81}, 123007 (2010)
  [arXiv:1004.1409 [astro-ph.CO]].

\bibitem{Maldacena:2002vr}
  J.~M.~Maldacena,
  JHEP {\bf 0305}, 013 (2003)
  [arXiv:astro-ph/0210603].



\bibitem{Lyth:2002my}
  D.~H.~Lyth, C.~Ungarelli and D.~Wands,
  Phys.\ Rev.\  D {\bf 67}, 023503 (2003)
  [arXiv:astro-ph/0208055].



\bibitem{Dimopoulos:2003ss}
  K.~Dimopoulos, G.~Lazarides, D.~Lyth and R.~Ruiz de Austri,
  Phys.\ Rev.\  D {\bf 68}, 123515 (2003)
  [arXiv:hep-ph/0308015].



\bibitem{Bartolo:2003jx}
  N.~Bartolo, S.~Matarrese and A.~Riotto,
  Phys.\ Rev.\  D {\bf 69}, 043503 (2004)
  [arXiv:hep-ph/0309033].


\bibitem{Lyth:2005fi}
 D.~H.~Lyth and Y.~Rodriguez,
 Phys.\ Rev.\ Lett.\  {\bf 95}, 121302 (2005)
 [arXiv:astro-ph/0504045].



\bibitem{Enqvist:2005pg}
  K.~Enqvist and S.~Nurmi,
  JCAP {\bf 0510}, 013 (2005)
  [arXiv:astro-ph/0508573].


\bibitem{Malik:2006pm}
  K.~A.~Malik and D.~H.~Lyth,
  JCAP {\bf 0609}, 008 (2006)
  [arXiv:astro-ph/0604387].

\bibitem{Sasaki:2006kq}
  M.~Sasaki, J.~Valiviita and D.~Wands,
  Phys.\ Rev.\  D {\bf 74}, 103003 (2006)
  [arXiv:astro-ph/0607627].

\bibitem{Huang:2008ze}
  Q.~G.~Huang,
  arXiv:0801.0467 [hep-th].

\bibitem{Ichikawa:2008iq}
  K.~Ichikawa, T.~Suyama, T.~Takahashi and M.~Yamaguchi,
  Phys.\ Rev.\  D {\bf 78}, 023513 (2008)
  [arXiv:0802.4138 [astro-ph]].

\bibitem{Multamaki:2008yv}
  T.~Multamaki, J.~Sainio and I.~Vilja,
  Phys.\ Rev.\  D {\bf 79}, 103516 (2009)
  [arXiv:0803.2637 [astro-ph]].


\bibitem{Enqvist:2008gk}
  K.~Enqvist and T.~Takahashi,
  JCAP {\bf 0809}, 012 (2008)
  [arXiv:0807.3069 [astro-ph]].

\bibitem{Huang:2008bg}
  Q.~G.~Huang and Y.~Wang,
  JCAP {\bf 0809}, 025 (2008)
  [arXiv:0808.1168 [hep-th]].

\bibitem{Huang:2008zj}
  Q.~G.~Huang,
  JCAP {\bf 0811}, 005 (2008)
  [arXiv:0808.1793 [hep-th]].


\bibitem{Kawasaki:2008mc}
  M.~Kawasaki, K.~Nakayama and F.~Takahashi,
  JCAP {\bf 0901}, 026 (2009)
  [arXiv:0810.1585 [hep-ph]].


\bibitem{Chingangbam:2009xi}
  P.~Chingangbam and Q.~G.~Huang,
  JCAP {\bf 0904}, 031 (2009)
  [arXiv:0902.2619 [astro-ph.CO]].


\bibitem{Enqvist:2009zf}
  K.~Enqvist, S.~Nurmi, G.~Rigopoulos, O.~Taanila and T.~Takahashi,
  JCAP {\bf 0911}, 003 (2009)
  [arXiv:0906.3126 [astro-ph.CO]].


\bibitem{Enqvist:2009eq}
  K.~Enqvist and T.~Takahashi,
  JCAP {\bf 0912}, 001 (2009)
  [arXiv:0909.5362 [astro-ph.CO]].

\bibitem{Enqvist:2009ww}
  K.~Enqvist, S.~Nurmi, O.~Taanila and T.~Takahashi,
  JCAP {\bf 1004}, 009 (2010)
  [arXiv:0912.4657 [astro-ph.CO]].

\bibitem{Chingangbam:2010xn}
  P.~Chingangbam and Q.~G.~Huang,
  arXiv:1006.4006 [astro-ph.CO].

\bibitem{Sefusatti:2009xu}
  E.~Sefusatti, M.~Liguori, A.~P.~S.~Yadav, M.~G.~Jackson and E.~Pajer,
  JCAP {\bf 0912}, 022 (2009)
  [arXiv:0906.0232 [astro-ph.CO]].

\bibitem{Byrnes:2008zy}
  C.~T.~Byrnes, K.~Y.~Choi and L.~M.~H.~Hall,
  JCAP {\bf 0902}, 017 (2009)
  [arXiv:0812.0807 [astro-ph]].

\bibitem{Kumar:2009ge}
  J.~Kumar, L.~Leblond and A.~Rajaraman,
  JCAP {\bf 1004}, 024 (2010)
  [arXiv:0909.2040 [astro-ph.CO]].

\bibitem{Byrnes:2009pe}
  C.~T.~Byrnes, S.~Nurmi, G.~Tasinato and D.~Wands,
  JCAP {\bf 1002}, 034 (2010)
  [arXiv:0911.2780 [astro-ph.CO]].

\bibitem{Byrnes:2010ft}
  C.~T.~Byrnes, M.~Gerstenlauer, S.~Nurmi, G.~Tasinato and D.~Wands,
  arXiv:1007.4277 [astro-ph.CO].

\bibitem{Huang:2010cy}
 Q.~G.~Huang,
 JCAP {\bf 1011}, 026 (2010).
[arXiv:1008.2641v4 [astro-ph.CO].]

\bibitem{Chen:2005fe}
  X.~Chen,
  Phys.\ Rev.\  D {\bf 72}, 123518 (2005)
  [arXiv:astro-ph/0507053].

\bibitem{LoVerde:2007ri}
  M.~LoVerde, A.~Miller, S.~Shandera and L.~Verde,
  JCAP {\bf 0804}, 014 (2008)
  [arXiv:0711.4126 [astro-ph]].

\bibitem{Khoury:2008wj}
  J.~Khoury and F.~Piazza,
  JCAP {\bf 0907}, 026 (2009)
  [arXiv:0811.3633 [hep-th]].

\bibitem{RenauxPetel:2009sj}
  S.~Renaux-Petel,
  JCAP {\bf 0910}, 012 (2009)
  [arXiv:0907.2476 [hep-th]].

\bibitem{Moroi:2002rd}
  T.~Moroi and T.~Takahashi,
  Phys.\ Rev.\  D {\bf 66}, 063501 (2002)
  [arXiv:hep-ph/0206026].

\bibitem{Lyth:2003ip}
  D.~H.~Lyth and D.~Wands,
  Phys.\ Rev.\  D {\bf 68}, 103516 (2003)
  [arXiv:astro-ph/0306500].

\bibitem{beltran:2008ei}
  M.~Beltran,
  Phys.\ Rev.\  D {\bf 78}, 023530 (2008)
  [arXiv:0804.1097 [astro-ph]].


\bibitem{Moroi:2008nn}
  T.~Moroi and T.~Takahashi,
  Phys.\ Lett.\  B {\bf 671}, 339 (2009)
  [arXiv:0810.0189 [hep-ph]].


\bibitem{Takahashi:2009cx}
  T.~Takahashi, M.~Yamaguchi and S.~Yokoyama,
  Phys.\ Rev.\  D {\bf 80}, 063524 (2009)
  [arXiv:0907.3052 [astro-ph.CO]].

\bibitem{Seery:2006js}
  D.~Seery and J.~E.~Lidsey,
  JCAP {\bf 0701}, 008 (2007)
  [arXiv:astro-ph/0611034].

\bibitem{Byrnes:2006vq}
  C.~T.~Byrnes, M.~Sasaki and D.~Wands,
  Phys.\ Rev.\  D {\bf 74}, 123519 (2006)
  [arXiv:astro-ph/0611075].


\end{thebibliography}
\end{document}